\documentclass[
  ]
{article}

\usepackage{graphicx}
\usepackage{url}             % For breaking URLs easily trough lines
\begin{document}

\title{On The Possible Mechanism Of Energy Dissipation In Shock-Wave Fronts Driven Ahead Of Coronal Mass Ejections
\thanks{Accepted for publication in Proceedings of the Solar Wind 12 Conference, Saint Malo, France, 21-26 June 2009}}

%\classification{96.60.ph, 96.50.Fm} \keywords{coronal mass
%ejection, shock wave}

\author{M.V. Eselevich and V.G. Eselevich\\
\it Institute of Solar-Terrestrial physics, Irkutsk, Russia}

%{  address={Institute of Solar-Terrestrial physics, Irkutsk,
%Russia} }

%\author{V.G. Eselevich}
%{
%  address={Institute of Solar-Terrestrial physics, Irkutsk, Russia}
%}

\maketitle

\begin{abstract}
Analysis of Mark 4 and LASCO C2, C3 coronagraph data shows that,
at the distance $R \leq 6$ R$_\odot$ from the center of the Sun,
the thickness of a CME-generated shock-wave front ($\delta _F$)
may be of order of the proton mean free path. This means that the
energy dissipation mechanism in the shock front at these distances
is collisional. A new discontinuity (thickness $\delta _F^* \ll
\delta _F$) is observed to appear in the anterior part of the
front at $R \geq 10$ R$_\odot$. Within the limits of experimental
error, the thickness $\delta _F^* \approx$ 0.1-0.2 R$_\odot$ does
not vary with distance and is determined by the spatial resolution
of the LASCO C3 instrument. At the initial stage of formation, the
discontinuity on the scale of $\delta _F^*$ has rather small
amplitude and exists simultaneously with the front having
thickness $\delta _F$. The relative amplitude of the discontinuity
gradually increases with distance, and the brightness profile
behind it becomes even. Such transformations may be associated
with the transition from a collisional shock wave to a
collisionless one.
\end{abstract}

%\maketitle

%%%%%%%%%%%%%%%%%%%%%%%%%%%%%%%%%%%%%%%%%%%%
%% MAINMATTER
%%%%%%%%%%%%%%%%%%%%%%%%%%%%%%%%%%%%%%%%%%%%

\section{Introduction}
Eselevich M. and V. \cite{Esel2008} revealed a disturbed region
extending along the direction of propagation ahead of a coronal
mass ejection (CME) when its velocity, $u$, was below a certain
critical velocity, $u_C$, relatively to the ambient coronal
plasma. Given $u > u_C$, a shock wave with front thickness $\delta
_F$ was formed in the frontal part of the disturbed region. In
\cite{Esel2009} also was shown that the coordinate system
connected with the CME's frontal structure was best suited for
demonstrating differences of the disturbed region, reflecting the
presence or absence of a shock wave. Moreover, the possibility of
relatively accurate measurements of $\delta _F$ in the solar
corona with Mark 4 and LASCO C2 was justified. Purpose of this
work is to analyze a possible dissipation mechanism in the shock
front, using measurements of the shock wave thickness.

\section{Method of analysis}
For this study, we analysed coronal images obtained by LASCO C2
and C3 onboard the SOHO spacecraft \cite{Brue1995}, presented as
difference brightness $\Delta P = P(t) - P(t_0)$, where $P(t_0)$
is the undisturbed brightness at a moment $t_0$, before the event
considered; $P(t)$ is the disturbed brightness at $t > t_0$. We
used calibrated LASCO images with the total brightness $P(t)$
expressed in terms of the mean solar brightness (P$_{msb}$).

For 1.2 R$_\odot < R <$ 2 R$_\odot$, we used polarization
brightness images from the ground-based coronagraph-polarimeter
Mark 4 (Mauna Loa Solar Observatory,
\url{http://mlso.hao.ucar.edu}). As was the case with LASCO data,
these images were expressed in terms of difference brightness.

\section{Identification of shock front ahead of a CME}
A shock front ahead of a CME can be identified reliably only in
the most simple cases. Particularly when:
\begin{enumerate}
\item A CME has a three-part structure and consists of a frontal
structure (FS), cavity, and bright core (sometimes the core can be
absent).
\item A CME propagates near the plane of the sky. This
means that the measured CME velocity is close to the true radial
velocity.
\item Registration of shock wave front is implemented in
the finite area in the direction of the CME motion.
\end{enumerate}

In order to show that the observed discontinuity in the brightness
distribution is a shock-wave front and not a current sheet ahead
of, but associated with the CME, we used two different
complementary approaches:
\begin{enumerate}
\item The disturbed region state was investigated for some CMEs
with different velocities, $u$. For CMEs with velocities $u >
u_C$, a shock front was detected in the frontal part of the
disturbed region ($u_C$ is the critical velocity that is about the
local Alfven velocity in the corona).
\item The evolution process of the disturbed region and shock
wave formation was studied for specified CMEs, in the coordinate
system connected with the frontal structure, as the CME velocity
$u$ overcomes the critical velocity $u_C$ \cite{Esel2009}.
\end{enumerate}

Within the limits of these two approaches, eight CMEs with
velocities in the range 700-2500 km s$^{-1}$ were examined.

One of the most important parameters of the shock front is its
thickness $\delta _F$, since it contains information about the
energy dissipation mechanism in the shock wave. Analysis in
\cite{Esel2008,Esel2009} showed that it is possible to measure the
shock front thickness $\delta _F$ correctly with Mark 4 and LASCO
C2 data. Due to measurement results, an experimental dependence
$\delta _F(R)$ was constructed for eight CMEs with velocities $u >
u_C$ (upper curve made-up of symbols labelled as $\delta _F$ in
Figure~1). The solid thin line in Figure~1 denotes average curve
for $\delta _F$.

\begin{figure}%[!b]
\begin{center}
  \includegraphics[width=.8\textwidth]{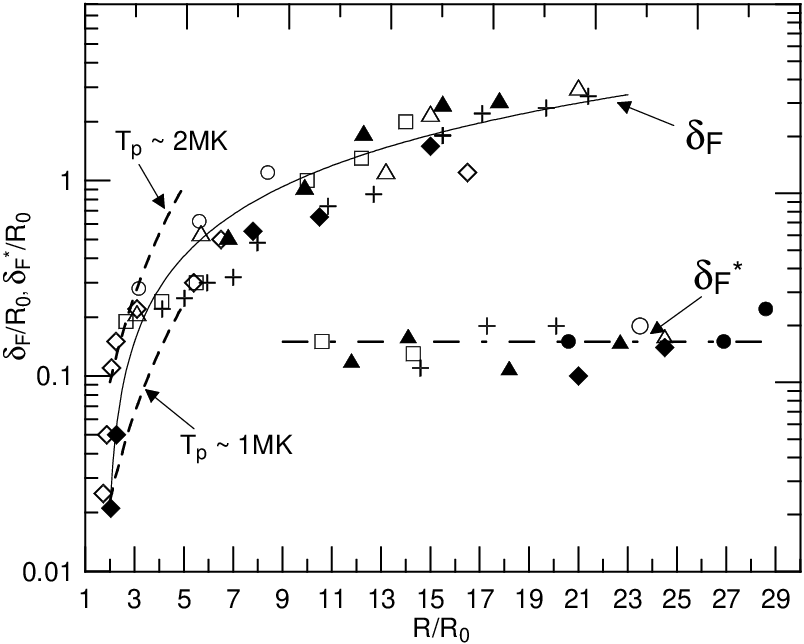}
\end{center}
  \caption{Variation in the front thickness of collisional $\delta _F$ (thin solid curve)
  and collisionless $\delta _F^*$ (dash-and-dot line)
  CME-generated shock waves with distance $R$ from the solar center,
  in eight CMEs with high velocities: solid circles denote 11 June 1998,
  $PA =$ 80$^\circ$; crosses -- 3 March 2000, $PA =$ 230$^\circ$;
  solid triangles -- 28 June 2000, $PA =$ 270$^\circ$;
  empty squares -- 4 September 2000, $PA =$ 300$^\circ$;
  empty triangles -- 21 April 2002, $PA =$ 270$^\circ$;
  empty circles -- 4 November 2003, $PA =$ 238$^\circ$;
  solid diamonds -- 22 November 2001, $PA =$ 247-254$^\circ$;
  empty diamonds -- 26 October 2003, $PA =$ 265-290$^\circ$
  (from Mark 4 and LASCO C2, C3 data).
  The heavy-dashed curves are the proton mean free path $\lambda _p$ calculated for two proton temperatures
  ($T_p = 10^6$ K and $T_p = 2\times 10^6$ K).}
\end{figure}

As is evident from the plot, the registered shock front thickness
$\delta _F$ increases with distance. The most significant
variation in $\delta _F$ is observed near the Sun. As an example,
Figure~2 shows three difference brightness profiles, constructed
with Mark 4 data, for the CME observed on 26 October 2003. At
distances from $\approx 1.7$ R$_\odot$ to $\approx 2$ R$_\odot$,
the shock front thickness increased fivefold (shock front is
cross-hatched in Figure~2). (Empty circles denote difference
brightness distribution before appearance of CMEs and can serve as
an estimate for the noise level).

\begin{figure}
\begin{center}
  \includegraphics[width=.8\columnwidth]{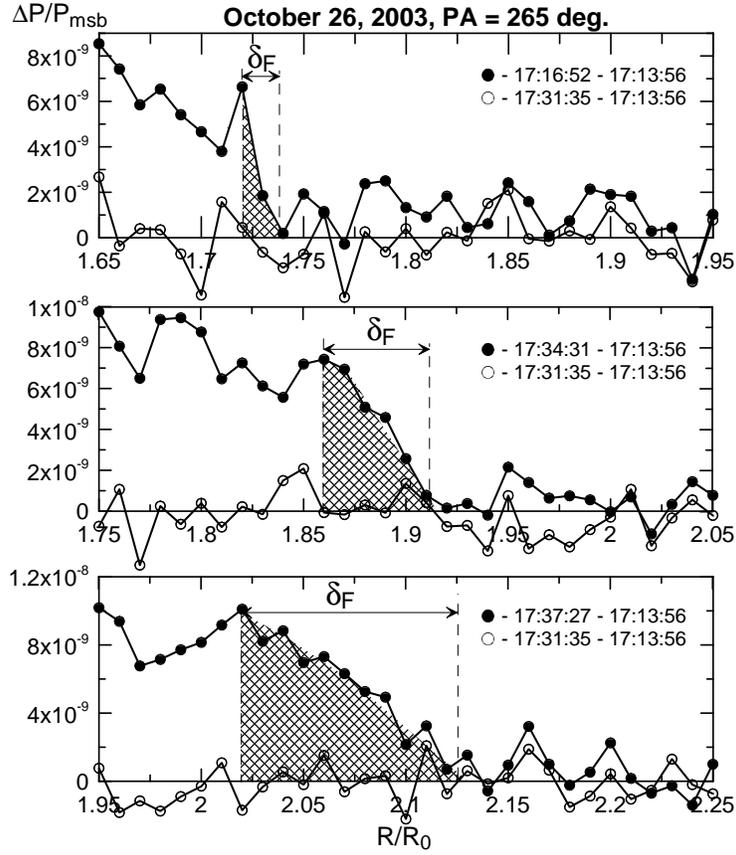}
\end{center}
  \caption{CME on 26 October 1997. Distributions of the difference
  polarization brightness depending on distance, $R$, along the direction of
  the CME propagation ($PA\approx 265^\circ$) for three moments in time (from Mark 4 data).
  Empty circles denote the distribution of the difference brightness, just before the appearance of the CME ($t$ = 17:31:35).}
\end{figure}

Let us compare the observed shock wave front thickness $\delta _F$
with the proton mean free path, $\lambda _p$, in the corona. The
upper and lower dashed lines in Figure~1 correspond to the mean
free path computed using the formula $\lambda _p$/R$_\odot \approx
 10^{-7} T^2/N$ \cite[p. 14]{Schu2005}, respectively, for
temperatures $T_p \sim 10^6$ K and $T_p \sim 2\times 10^6$ K, and
the electron density profile from \cite{Stra2002}. The
characteristic front size is comparable with the free path
($\delta _F \sim \lambda _p$), at least up to $\sim 6$ R$_\odot$.
This implies that the dissipation mechanism in the shock wave may
be collisional at these distances.

Thus, we appear to encounter a rare situation where we can resolve
and examine the collisional shock front structure in the plasma.
This has not been possible so far, either in gas or in plasma,
because of the very small size of $\lambda _p$. This may be
regarded as the first experimental evidence of the theoretical
conclusion that the collisional shock wave front thickness is of
order of the proton mean free path \cite{Zeld1966}.

\section{On the possible transition from a collisional shock wave to a collisionless
one}

The free mean path, and, consequently, front thickness $\delta _F$
of a collisional shock wave increase away from the Sun. At a
distance of more than 6 R$_\odot$, the shock front structure must
be eventually rearranged and transformed to a collisionless shock
wave with the front thickness $\delta _F^* \ll \lambda _p$. This
is evident from the fact that, in interplanetary space and at the
Earth's orbit in particular, collisionless shock waves are
observed ahead of ICMEs. The possibility of such a transformation
has not been studied in the past either theoretically or
experimentally.

Formation of a small-scale discontinuity with thickness $\delta
_F^* \ll \delta _F, \lambda _p$ in the brightness distribution may
be considered as a feature of a collisionless shock wave front.
Let us examine the dynamics of the shock front for the CME of 20
September 1997 at distances $R > 6$ R$_\odot$ as an example.
Figure~3 shows difference brightness distributions $\Delta P(t,
R)$ at successive moments of time plotted along the direction of
the CME propagation (range of position angles $€ =$
270-280$^\circ$. The collisional shock front is depicted by the
crosshatching in plots. Its thickness $\delta _F$ increases with
distance in accordance with dependence $\delta _F(R)$ from
Figure~1. Formation of a new discontinuity is observed in the
anterior part of the front, from $R\approx 20$ R$_\odot$ and
onward. At $R\approx 23$ R$_\odot$, the thickness of the
discontinuity is $\delta _F^*\approx 0.15$ R$_\odot$ (two lower
panels in Figure~3).

\begin{figure}
\begin{center}
  \includegraphics[width=.7\columnwidth]{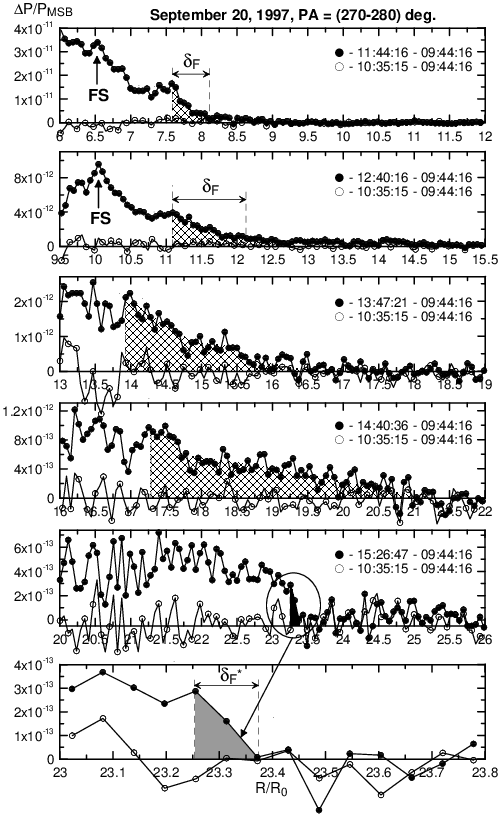}
\end{center}
  \caption{CME on 20 September 1997. Difference brightness
  distributions depending on distance $R$ along the direction of
  the CME propagation ($PA\approx$ 270-280$^\circ$) at successive
  moments of time (from LASCO C3 data). Empty circles denote
  the distribution of the difference brightness just before the appearance of the CME ($t$ = 10:35:15).}
\end{figure}

Formation of the discontinuity with thickness $\delta _F^* \ll
\lambda _p$ at 10 R$_\odot \leq R \leq 30$ R$_\odot$ was observed
for all eight CMEs used for constructing  the plot of $\delta
_F(R)$ in Figure~1. Within the limits of experimental error,
$\delta _F^*$ is about 0.1-0.2 R$_\odot$ and is independent of
distance (horizontal dash-and-dot line in Figure~1). At the
initial stage of formation, the discontinuity has a rather small
amplitude and exists simultaneously with the collisional front
having thickness $\delta _F$. The relative amplitude of the
discontinuity gradually increases with distance, and the
brightness profile behind it becomes even. Thus, transition occurs
from the front with thickness $\delta _F$ to the discontinuity
with thickness $\delta _F^* \ll \delta _F$.

The spatial resolution of the C3 instrument is $K\approx 0.12$
R$_\odot$ and is approximately equal to the observed discontinuity
thickness $\delta _F^*$. This means that the real discontinuity
scale can be much less than the thickness being observed. This,
together with the fact that $\delta _F^*$ remains constant with
distance, implies that this discontinuity is a collisionless shock
wave whose observed front thickness is unresolved and determined
by the spatial resolution of the C3 coronagraph. Notice that
similar discontinuities in the brightness profiles were registered
ahead of fast ($V > 1500$ km s$^{-1}$) halo-type CMEs at distances
of more than 10 R$_\odot$ in \cite{Onti2009}. The authors also
associated these discontinuities with collisionless shock waves.

It is notable that the thickness $\delta _F$ of the collisional
shock front measured at the minimum distance $R \approx 1.7$
R$_\odot$ from the Sun is $\approx 0.015$ R$_\odot$ (see Figures~1
and 2). This is an order of magnitude less than the thickness
$\delta _F^* \approx 0.15$ R$_\odot$ of the collisionless front,
determined by spatial resolution of the C3 instrument.

\section{Conclusions}
Mark 4 and LASCO C2, C3 coronagraph data analysis shows that, at
the distance $R\leq 6$ R$_\odot$ from the Sun center along the
streamer belt, the thickness $\delta_ F$ of the CME-generated
shock front may be of the order of the proton mean free path. This
means that the energy dissipation mechanism in the shock front at
these distances is collisional. A new discontinuity with thickness
$\delta _F^* \ll \delta _F$ is observed to appear in the anterior
part of the front at $R \geq 10$ R$_\odot$. Within the limits of
experimental error, the thickness $\delta _F^* \approx$ 0.1-0.2
R$_\odot$ does not vary with distance and is determined by the
spatial resolution of the LASCO C3 instrument. At the initial
stage of formation, the discontinuity has a rather small amplitude
and exists simultaneously with the front having a thickness
$\delta _F$. The relative amplitude of this discontinuity
gradually increases with distance, and the brightness profile
behind it becomes even. Such transformation may be associated with
the transition from a collisional shock wave to a collisonless
one.

%%%%%%%%%%%%%%%%%%%%%%%%%%%%%%%%%%%%%%%%%%%%
%% Sample figure:
%%
%% The option [height=...] scales the picture to the given height,
%% without it it would be printed at its nominal size
%%%%%%%%%%%%%%%%%%%%%%%%%%%%%%%%%%%%%%%%%%%%

%\begin{figure}
%  \includegraphics[height=.3\textheight]{golfer}
%  \caption{Picture to fixed height}
%\end{figure}

%%%%%%%%%%%%%%%%%%%%%%%%%%%%%%%%%%%%%%%%%%%%%%%%
%% BACKMATTER
%%%%%%%%%%%%%%%%%%%%%%%%%%%%%%%%%%%%%%%%%%%%%%%%

\subsection*{Acknowledgments} The work was supported by program No.
16 part 3 of the Presidium of the Russian Academy of Sciences,
program of state support for leading scientific schools
NS-2258.2008.2, and the Russian Foundation for Basic Research
(Project No. 09-02-00165a). The SOHO/LASCO data used here are
produced by a consortium of the Naval Research Laboratory (USA),
Max-Planck-Institut fuer Aeronomie (Germany), Laboratoire
d'Astronomie (France), and the University of Birmingham (UK). SOHO
is a project of international cooperation between ESA and NASA.
The Mark 4 data are courtesy of the High Altitude
Observatory/NCAR.

%%%%%%%%%%%%%%%%%%%%%%%%%%%%%%%%%%%%%%%%%%%%%%%%
%% The bibliography can be prepared using the BibTeX program or
%% manually.
%%
%% The code below assumes that BibTeX is used.  If the bibliography is
%% produced without BibTeX comment out the following lines and see the
%% aipguide.pdf for further information.
%%
%% For your convenience a manually coded example is appended
%% after the \end{document}
%%%%%%%%%%%%%%%%%%%%%%%%%%%%%%%%%%%%%%%%%%%%%%%%

%%%%%%%%%%%%%%%%%%%%%%%%%%%%%%%%%%%%%%%%%%%%%%%%
%% You may have to change the BibTeX style below, depending on your
%% setup or preferences.
%%
%%
%% For The AIP proceedings layouts use either
%%%%%%%%%%%%%%%%%%%%%%%%%%%%%%%%%%%%%%%%%%%%

%\bibliographystyle{aipproc}   % if natbib is available
%\bibliographystyle{aipprocl} % if natbib is missing

%%%%%%%%%%%%%%%%%%%%%%%%%%%%%%%%%%%%%%%%%%%
%% You probably want to use your own bibtex database here
%%%%%%%%%%%%%%%%%%%%%%%%%%%%%%%%%%%%%%%%%%%
%\bibliography{sample}

%%%%%%%%%%%%%%%%%%%%%%%%%%%%%%%%%%%%%%%%%%%
%% Just a reminder that you may have to run bibtex
%% All of it up to \end{document} can be removed
%% if you don't like the warning.
%%%%%%%%%%%%%%%%%%%%%%%%%%%%%%%%%%%%%%%%%%%
%\IfFileExists{\jobname.bbl}{}
% {\typeout{}
%  \typeout{******************************************}
%  \typeout{** Please run "bibtex \jobname" to optain}
%  \typeout{** the bibliography and then re-run LaTeX}
%  \typeout{** twice to fix the references!}
%  \typeout{******************************************}
%  \typeout{}
% }

\end{document}